\documentclass{amsart}
\usepackage{amssymb,latexsym}
\usepackage{graphicx}

\newcommand{\de}{\mathrm{d}}
\newcommand{\e}{\mathrm{e}}
\newcommand{\nn}{\nonumber}
\newcommand{\ug}{\!\!\!\!&=&\!\!\!\!}

\begin{document}

\title[On integrals involving Hermite polynomials]{On integrals involving Hermite polynomials}
\author{D. Babusci$^\dag$, G. Dattoli$^\ddag$, M. Quattromini$^\ddag$}
\address{$^\dag$ INFN - Laboratori Nazionali di Frascati, via E. Fermi 40, I-00044 Frascati.}
\address{$^\ddag$ ENEA - Centro Ricerche Frascati, via E. Fermi 45, I-00044 Frascati.}

\begin{abstract}
We show how the combined use of the generating function method and of the theory of multivariable 
Hermite polynomials is naturally suited to evaluate integrals of gaussian functions and of multiple 
products of Hermite polynomials.\end{abstract}

\maketitle

The generating function method is often exploited to derive the analytic form of integrals of the type \cite{Book}
\begin{equation}\label{integr}
I_n = \int_{- \infty}^\infty \de x\, H_n (a\,x + b, y)\e^{\,- c\,x^2 + \alpha\,x}
\end{equation}
where
\begin{equation}
H_n (x,y) = n!\,\sum_{k = 0}^{[n/2]} \frac{x^{n - 2\,k}\,y^k}{(n - 2\,k)!\, k!}
\end{equation}
is the two variable Hermite polynomial with generating function \cite{Appe}
\begin{equation}\label{genfun}
\sum_{n = 0}^\infty \frac{t^n}{n!}\,H_n (x, y) = \e^{x\,t + y\,t^2}\,.
\end{equation}
By taking into account this identity, from eq. \eqref{integr} we obtain
\begin{equation}
\sum_{n = 0}^\infty \frac{t^n}{n!}\,I_n = \e^{b\,t + y\,t^2}\,\int_{- \infty}^\infty \de x\,\e^{\,- c\,x^2 + (a\,t + \alpha)\,x}\,,
\end{equation}
and, thus, we have reduced our problem to the evaluation of a trivial gaussian integral, which yields 
\begin{equation}
\sum_{n = 0}^\infty \frac{t^n}{n!}\,I_n = \frac{\sqrt{\pi}}{\sqrt{c}}\,\exp\left\{\frac{\alpha^2}{4\,c} + 
\left(b + \frac{\alpha\,a}{2\,c}\right)\,t + \left(y + \frac{a^2}{4\,c}\right)\,t^2\right\}\,.
\end{equation}

The use of the generating function \eqref{genfun} allows us to write the integral \eqref{integr} as follows
\begin{equation}
I_n = \frac{\sqrt{\pi}}{\sqrt{c}}\,\exp\left(\frac{\alpha^2}{4\,c}\right)\,
H_n \left(b + \frac{\alpha\,a}{2\,c}, y + \frac{a^2}{4\,c}\right)\,.
\end{equation}
Moreover, since
\begin{equation}
H_n (x, y) = \exp\left(\,y\,\partial_x^2\right)\,x^n\,, 
\end{equation}
we can write $I_n$ in the following operational form 
\begin{equation}
\label{eq:intIn}
I_n =  \frac{\sqrt{\pi}}{\sqrt{c}}\,\exp\left(\frac{\alpha^2}{4\,c}\right)\,
\exp\left\{\left(y + \frac{a^2}{4\,c}\right)\,\partial_b^2 + \frac{\alpha\,a}{2\,c}\,\partial_b\right\}\,b^n\,.
\end{equation}

\vspace{1.0cm}
Let us now consider the integral
\begin{equation}
{}_mI_n = \int_{- \infty}^\infty \de x\,x^m\,H_n (a\,x + b, y)\e^{\,- c\,x^2 + \alpha\,x}\,,
\end{equation}
that occurs in some problems involving quantum harmonic oscillator, or, in classical optics, in the evaluation of 
overlapping of Gauss-Hermite beams. It can be cast in the form
\begin{equation}\label{deriv}
{}_mI_n = \partial_\alpha^m\,I_n\,,
\end{equation}
from which, taking into account the identities
\begin{equation}
\partial_x^k\,H_n (x, y) = \frac{n!}{(n - k)!}\,H_{n - k} (x, y) \qquad \partial_x^k\,\e^{\,\beta\,x^2} = 
H_k (2\,\beta\,x, \beta)\,\e^{\,\beta\,x^2}\,,
\end{equation}
one finds 
\begin{equation}
{}_mI_n = \frac{\sqrt{\pi}}{\sqrt{c}}\,\exp\left(\frac{\alpha^2}{4\,c}\right)\,
H_{m,n} \left(\frac{\alpha}{2\,c}, \frac1{4\,c}; b + \frac{\alpha\,a}{2\,c}, y + \frac{a^2}{4\,c} \mid \frac{a}{2\,c}\right)\,,
\end{equation}
where we have introduced the two-index Hermite polynomials \cite{Appe,Datto} defined as follows 
\begin{equation}
H_{m,n}  (x, y; w, z \mid \tau) = \sum_{k = 0}^{\mathrm{min} (m,n)} \frac{m!\,n!}{(m - k)!\,(n - k)!\,k!}\,
\tau^k\,H_{m - k} (x, y)\,H_{n - k} (w, z)\,.
\end{equation}
The relevant generating function writes
\begin{equation}
\sum_{m, n = 0}^\infty \frac{u^m}{m!}\,\frac{v^n}{n!}\,H_{m,n} (x, y; w, z \mid \tau) = 
\exp\left(x\,u + y\,u^2 + v\,w + z\,v^2 + \tau\,u\,v\right)\,,
\end{equation}
and, thus, using the same procedure as before, we find for integrals of the type
\begin{equation}
I_{m,n} = \int_{- \infty}^\infty \de x\,H_m (a\,x + b, y)\,H_n (c\,x + d, z)\,\e^{\,- f\,x^2 + \alpha\,x}
\end{equation}
the explicit form 
\begin{equation}
\label{eq:Imnint}
I_{m,n} = \frac{\sqrt{\pi}}{\sqrt{f}}\,\exp\left(\frac{\alpha^2}{4\,f}\right)\,
H_{m,n} \left(\bar{x}, \bar{y}; \bar{w}, \bar{z} \mid \tau \right)\,,
\end{equation}
where we put
\begin{equation}
\bar{x} = b + \frac{a}{2\,f}\,\alpha\,, \quad \bar{y} = y + \frac{a^2}{4\,f}\,, \quad 
\bar{w} = d + \frac{c}{2\,f}\,\alpha\,, \quad \bar{z} = z + \frac{c^2}{4\,f}\,, \quad 
\tau = \frac{a\,c}{2\,f}\,. \nn
\end{equation}
By taking into account that
\begin{equation}
H_{m,n} \left(x, y; w, z \mid \tau \right) = \exp\left(y\,\partial_x^2 + w\,\partial_z^2 + \tau\,\partial_{xz}^2\right)\,x^m\,y^n
\end{equation}
we obtain the generalization of eq. (\ref{eq:intIn}) as
\begin{equation}
I_{m,n} = \frac{\sqrt{\pi}}{\sqrt{f}}\,\exp\left(\frac{\alpha^2}{4\,f}\right)\,\exp\left(\bar{y}\,\partial_{\bar{x}}^2 + 
\bar{w}\,\partial_{\bar{z}}^2 + \tau\,\partial_{\bar{x}\bar{z}}^2\right)\,\bar{x}^m\,\bar{y}^n\,.
\end{equation}

The operatorial method can be applied from the very beginning and the integral (\ref{eq:Imnint}) can be written as
\begin{equation}
I_{m,n} = \exp\left(y\,\partial_b^2 + z\,\partial_d^2\right)\,\mathcal{I}_{m,n}\,,
\end{equation}
where
\begin{eqnarray}
\mathcal{I}_{m,n} \ug \int_{- \infty}^\infty \de x\,(a\,x + b)^m\,(c\,x + d)^n\,\e^{\,- f\,x^2 + \alpha\,x} \nn \\
\ug \frac{\sqrt{\pi}}{\sqrt{f}}\,\exp\left(\frac{\alpha^2}{4\,f}\right)\,H_{m,n} \left(\bar{x}, \frac{a^2}{4\,f}; \bar{w}, \frac{c^2}{4\,f} \mid \tau \right)\,.
\end{eqnarray}
This approach may simplify the computation of integrals involving products of more than two Hermite polynomials. 

It can now be shown that for the integral
\begin{equation}\label{intpmn}
{}_pI_{m,n} = \int_{- \infty}^\infty \de x\,x^p\,H_m (a\,x + b, y)\,H_n (c\,x + d, z)\,\e^{\,- f\,x^2 + g\,x}\,,
\end{equation}
an identity analogous to the \eqref{deriv} holds
\begin{equation}
{}_pI_{m,n} = \partial_\alpha^p\,I_{m,n}\,.
\end{equation}
The use of the identities 
\begin{eqnarray}
\partial_x^k\,H_{m,n}  (x, y; w, z \mid \tau) \ug \frac{m!}{(m - k)!}\,H_{m - k,n}  (x, y; w, z \mid \tau) \nn \\
\partial_w^k\,H_{m,n}  (x, y; w, z \mid \tau) \ug \frac{n!}{(n - k)!}\,H_{m,n - k}  (x, y; w, z \mid \tau)\,,
\end{eqnarray}
allows us to obtain for the integral \eqref{intpmn} the following expression 
\begin{equation}
{}_pI_{m,n} = \frac{\sqrt{\pi}}{\sqrt{f}}\,\exp\left(\frac{\alpha^2}{4\,f}\right)\,
\sum_{k = 0}^p \binom{p}{k}\,H_{p - k} \left(\frac{\alpha}{2\,f}, \frac1{4\,f}\right)\,R_{m,n}^{(k)}
\end{equation} 
where
\begin{eqnarray}
R_{m,n}^{(k)} \ug \partial_\alpha^k\,H_{m,n} \left(\bar{x}, \bar{y}; \bar{w}, \bar{z} \mid \tau \right) \nn \\
\ug \left(\frac{c}{2\,f}\right)^k\, \sum_{l = 0}^k \binom{k}{l} \left(\frac{a}{c}\right)^k 
\frac{m!\,n!}{(m - l)!\,(n - k + l)!}\,H_{m - j,n -k + l} \left(\bar{x}, \bar{y}; \bar{w}, \bar{z} \mid \tau \right)\,.
\end{eqnarray}

\vspace{1.0cm}
As a final example, we consider the integral
\begin{equation}
I_n = \int_{- \infty}^\infty \de x\,\frac{(a\,x + b)^n}{(1 + c\,x^2)^\nu} \qquad\qquad 
(n < 2\,\nu, \nu \in \mathbb{R})\,.
\end{equation}
The combined use of the generating function and Laplace transform methods yields\footnote{In performing the sum we do not take 
into account of the condition $n < 2\,\nu$, but the consequent result in eq. (\ref{eq:Inint}) must be considered valid only under this hyphotesis.}
\begin{equation}
\sum_{n = 0}^\infty \frac{t^n}{n!}\,I_n = \frac{\e^{\,b\,t}}{\Gamma (\nu)}\,\int_0^\infty \de s\,e^{\,- s}\,s^{\nu - 1}\,
\int_{- \infty}^\infty \de x\,\e^{\,- s\,c\,x^2 + a\,x\,t}
\end{equation}
which can be finally explicitly worked out as
\begin{equation}
\label{eq:Inint}
I_n = \frac{\sqrt{\pi}}{\sqrt{c}\,\Gamma (\nu)}\,H_n^{(\nu)} \left(b, \frac{a^2}{4\,c}\right)\,,
\end{equation}
where
\begin{equation}
H_n^{(\nu)} (x, y) = n!\,\sum_{k = 0}^{[n/2]} \frac{\Gamma (\nu - k - 1/2)}{k!\,(n - 2\,k)!}\,x^{n - 2\,k}\,y^k\,.
\end{equation}
These polynomials do not belong to the Appell family and the relevant generating function can be written adopting an umbral notation as follows 
\begin{equation}
\sum_{n = 0}^\infty \frac{t^n}{n!}\,H_n^{(\nu)} (x, y) = \e^{\,x\,t + \hat{y}\,t^2} \qquad \hat{y}^k = \Gamma(\nu - k - 1/2)\,y^k\,. 
\end{equation}
With this assumption we also get
\begin{equation}
H_n^{(\nu)} (x, y) = \exp\left(\hat{y}\,\partial_x^2\right)\,x^n\,,
\end{equation} 
and then most of the properties of this family of polynomials can be derived straightforwardly.


\end{document}